\documentclass[letterpaper, 10 pt, conference]{ieeeconf}  

\IEEEoverridecommandlockouts                              

\usepackage{amsmath,amsthm,amssymb}
\usepackage{graphicx,fixltx2e}
\usepackage{hyperref}
\usepackage{xy}
\usepackage[]{algorithm2e}
\usepackage{bbm}
\usepackage{todonotes}
\usepackage{float}
\usepackage{cite}

\newcommand{\N}{\mathbb{N}}

\newcommand{\R}{\mathbb{R}}

\newcommand{\ptcnt}[1]{\stepcounter{#1}\arabic{#1}} 
\newcommand{\vol}{\text{vol}}
\newcommand{\mc}[1]{\mathcal{#1}}

\makeatletter
\newcommand{\pushright}[1]{\ifmeasuring@#1\else\omit\hfill$\displaystyle#1$\fi\ignorespaces}
\newcommand{\pushleft}[1]{\ifmeasuring@#1\else\omit$\displaystyle#1$\hfill\fi\ignorespaces}
\makeatother

\newenvironment{proposition}[2][Proposition]{\begin{trivlist}
\item[\hskip \labelsep {\bfseries #1}\hskip \labelsep {\bfseries #2.}]}{\end{trivlist}}

\allowdisplaybreaks

\title{\LARGE \bf
Controller Synthesis for Discrete-Time Polynomial Systems via Occupation Measures
}

\author{Weiqiao Han and Russ Tedrake
\thanks{Computer Science and Artificial Intelligence Laboratory, Massachusetts Institute of Technology, 77 Massachusetts Avenue, Cambridge, MA 02139, USA. {\tt\small weiqiaoh,russt@mit.edu}}%
}

\begin{document}

\newcounter{eqn}

\maketitle
\thispagestyle{empty}
\pagestyle{empty}

\begin{abstract}
In this paper, we design nonlinear state feedback controllers for discrete-time polynomial dynamical systems via the occupation measure approach.
We propose the discrete-time controlled Liouville equation, and use it to formulate the controller synthesis problem as an infinite-dimensional linear programming problem on measures, which is then relaxed as finite-dimensional semidefinite programming problems on moments of measures and their duals on sums-of-squares polynomials.
Nonlinear controllers can be extracted from the solutions to the relaxed problems.
The advantage of the occupation measure approach is that we solve convex problems instead of generally non-convex problems, and the computational complexity is polynomial in the state and input dimensions, and hence the approach is more scalable.
In addition, we show that the approach can be applied to over-approximating the backward reachable set of discrete-time autonomous polynomial systems and the controllable set of discrete-time polynomial systems under known state feedback control laws.
We illustrate our approach on several dynamical systems.

\end{abstract}

\section{Introduction}
Given a discrete-time polynomial dynamical system and a target set in state space, we are interested in designing controllers that steer the system to the target set without violating state or control input constraints.
Controller synthesis for polynomial systems is a challenging problem in robotics and control.
Traditional approaches include designing a linear quadratic regulator (LQR) based on linearized dynamics in a neighborhood of the fixed point, model predictive control (MPC), feedback linearization, dynamic programming, and Lyapunov-based approaches.
These approaches each have their limitations.
LQR control and linear MPC only work for a small region around the fixed point.
To plan for the entire state space, the LQR-Trees method \cite{tedrake2010lqr} and the approximate explicit-MPC method \cite{marcucci2017approximate} have been invented.
Feedback linearization does not work if there are limits on the inputs.
Dynamic programming only works for systems with small dimensionality.
Lyapunov-based approaches are generally non-convex, but can be convexified by incorporating the integrator into the controller structure \cite{md2013controller} or adding delayed states in the Lyapunov function \cite{pitarch2016control}.

Recently the area has seen the development of the occupation measure approach \cite{lasserre2008nonlinear} (also known as the Lasserre hierarchy strategy on occupation measures \cite{henrion2016lasserre}).
The general framework of the approach is to first formulate the problem as an infinite-dimensional LP on measures and its dual on continuous functions, and to then approximate the LP by a hierarchy of finite-dimensional semidefinite programming (SDP) programs on moments of measures and their duals on sums-of-squares (SOS) polynomials.
The earliest notable application of the approach is the outer approximation of the region of attraction of continuous-time polynomial systems \cite{henrion2014convex}. 
The advantage of the approach is that the problem is formulated as a series of convex optimization problems instead of general non-convex problems, and theoretically the approximation to the real set can be made arbitrarily close.
Since then the occupation measure approach has been attracting increasing attention and study.
It has been applied to the approximation of the region of attraction, the backward reachable set, and the maximum controllable set for continuous-time polynomial systems \cite{korda2013inner, korda2014controller, korda2014convex, shia2014convex}.
It has also been applied to controller synthesis for continuous-time nonhybrid/hybrid polynomial systems \cite{majumdar2014convex, korda2014controller,zhao2017optimal}.

Studies on discrete-time polynomial systems, however, are relatively sparse compared to those on continuous-time polynomial systems.
In \cite{savorgnan2009discrete}, the authors considered the discrete-time nonlinear stochastic optimal control problem, which can be interpreted in terms of the Bellman equation.
In \cite{magron2017semidefinite}, the authors proposed the discrete-time Liouville equation and used it to formulate an optimization problem that approximates the forward reachable set of discrete-time autonomous polynomial systems.

We are particularly interested in discrete-time systems.
One reason is that any physical system simulated by a digital computer is discrete in time, and the control input sent by the digital computer is also discrete in time.
When modeling robots making and breaking contact with the environment, the continuous-time systems using some contact models need to handle measure differential inclusions for impacts \cite{posa2016stability}, while the discrete-time models equally capture the complexity of the constrained hybrid dynamics without worrying about impulsive events and event detection \cite{han2017feedback, marcucci2017approximate}.
As another related example, in $N$-step capturability analysis used to study balancing in legged robots, the decision-making is discrete on a footstep-to-footstep level, and the entire problem formulation is asking about the viability kernel, also known as the backward reachable set \cite{posa2017balancing}.

In this paper, we propose a controller synthesis method for discrete-time polynomial systems via the occupation measure approach.
We propose the discrete-time controlled Liouville equation, and use it to formulate the problem as an infinite-dimensional LP, approximated by a family of finite-dimensional SDP's.
By solving SDP's of certain degrees, we are able to extract controllers as polynomials of the corresponding degrees.
Unlike Lyapunov-based approaches, our controller synthesis process does not simultaneously return the controllable region, and hence the stability of the closed-loop system has to be checked a posteriori.
Nevertheless, we show that our approach can be applied to over-approximating both the backward reachable set of discrete-time autonomous polynomial systems and the controllable set of discrete-time polynomial systems given any polynomial state feedback control law. 
We illustrate our approach on several dynamical systems.
Our work can be viewed as the discrete-time counterpart of \cite{majumdar2014convex}, and a pathway towards controller synthesis for discrete-time hybrid polynomial systems. 

\section{Problem formulation}
\subsection{Problem statement}
Let $n,m \in \N$.
Consider the discrete-time control-affine polynomial system
\[ x_{t+1} = \phi(x_t,u_t) :=  f(x_t) + g(x_t) u_t. \]
The sets $X \subseteq \R^n$ and $U \subseteq \R^m$ are state and control input constraint sets, respectively.
The vectors $x_t \in X$ and $u_t \in U$ represent states and control inputs at time $t\in \N$, respectively.
$f(x)$ and $g(x)$ are polynomial maps.
Denote the target set by $Z \subseteq X$. 
Our goal is to design a polynomial state feedback controller $u_t = u(x_t) \in U$ that steers the system to the target set $Z$ without violating state and control input constraints.

$\R[x]$ (resp. $\R[u]$) stands for the set of polynomials in the variable $x = (x_1,\ldots,x_n)$ (resp. $u = (u_1,\ldots,u_m)$).
$\R_{2r}[x]$ (resp. $\R_{2r}[u]$) stands for the set of polynomials in the variable $x$ (resp. $u$) of degree at most $2r$.

Assume 
\[ X := \{ x\in \R^n: h^X_i(x) \geq 0, h^X_i(x)\in \R[x],i=1,\ldots,n_X\}, \]
is a compact basic semi-algebraic set.
Furthermore, assume that the moments of the Lebesgue measure on $X$ are available.
For example, if $X$ is an $n$-dimensional ball or box, then it satisfies this assumption.

Assume 
\begin{align*}
    U &:= \{ u\in \R^m: h^U_i(u) \geq 0, h^U_i(u)\in \R[u],i=1,\ldots,n_U\}\\
      & = [a_1,b_1]\times \ldots \times [a_m,b_m],
\end{align*}
where $a_i,b_i \in \R,i=1,\ldots,m$ are upper and lower limits on control inputs.
Furthermore, without loss of generality, assume 
\[ U := [-1,1]^m,\]
because the dynamics equation can be scaled and shifted.

Assume 
\[ Z := \{ x\in \R^n: h^Z_{i}(x)\geq 0, h^Z_i(x)\in \R[x],i=1,\ldots,n_Z\}, \]
is a compact basic semi-algebraic set.
In practice, we may choose $Z$ to be a small ball or box around the origin.

\subsection{Notations}
In this subsection, we introduce some notations in real analysis, functional analysis, and polynomial optimization.
For an introduction to these three subjects, please refer to \cite{folland2013real}, \cite{conway2013course}, and \cite{lasserre2010moments}, respectively.

Let $X \subseteq \R^n$ be a compact set.
$\mathcal{C}(X)$ denotes the Banach space of continuous functions on $X$ equipped with the sup-norm.
Its topological dual, denoted by $\mathcal{C}'(X)$, is the set of all continuous linear functionals on $\mathcal{C}(X)$.
$\mathcal{M}(X)$ denotes the Banach space of finite signed Radon measures on the Borel $\sigma$-algebra $\mathcal{B}(X)$ equipped with the total variation norm.
By Riesz Representation Theorem, $\mathcal{M}(X)$ is isometrically isomorphic to $\mathcal{C}'(X)$.
$\mathcal{C}_+(X)$ (resp. $\mathcal{M}_+(X)$) denotes the cone of non-negative elements of $\mathcal{C}(X)$ (resp. $\mathcal{M}(X)$).
The topology in $\mathcal{C}_+(X)$ is the strong topology of uniform convergence while the topology in $\mathcal{M}_+(X)$ is the weak-star topology.
For any $A\in \mathcal{B}(X)$, $\lambda_A$ denotes the restriction of the Lebesgue measure on $A$.
For $\mu,\nu\in \mathcal{M}(X)$, we say $\mu$ is dominated by $\nu$, denoted by $\mu\leq \nu$, if $\nu-\mu \in \mathcal{M}_+(X)$.

Define $r^X_i := \lceil {\deg h^X_i / 2} \rceil, i = 1,\ldots,n_X$,
$r^U_i := \lceil {\deg h^U_i / 2} \rceil, i = 1,\ldots,n_U$,
and $r^Z_i := \lceil {\deg h^Z_i / 2} \rceil, i = 1,\ldots,n_Z$.
$\Sigma[x]$ (resp. $\Sigma_r[x]$) denotes the cone of SOS polynomials (resp. SOS polynomials of degree up to $2r$) in the variable $x$.
$\mathbf{Q}^X_r$ (resp. $\mathbf{Q}^{XU}_r$, $\mathbf{Q}_r^Z$) denotes the $r$-truncated quadratic module generated by the defining polynomials of $X$ (resp. $X\times U$, $Z$), assuming $h^X_0(x) = 1$ (resp. $h^U_0(u) = 1$, $h^Z_0(x) = 1$):
\begin{align*}
    &\mathbf{Q}^X_r := \big\{\sum_{i=0}^{n_X} \sigma_i(x) h^X_i(x): \sigma_i \in \Sigma_{r-r^X_i}[x], i=0,\ldots,n_X \big\},\\
    &\mathbf{Q}^{XU}_r := \big\{\sum_{i=0}^{n_X} \sigma^X_i(x,u) h^X_i(x) + \sum_{i=0}^{n_U} \sigma^U_i(x,u) h^U_i(u):\\ 
    &\quad \quad \quad \quad \quad \sigma^X_i \in \Sigma_{r-r^X_i}[x,u], \sigma^U_j \in \Sigma_{r-r^U_j}[x,u],\\
    &\quad \quad \quad \quad \quad  i=0,\ldots,n_X, j=0,\ldots,n_U  \big\},\\
    &\mathbf{Q}^Z_r := \big\{\sum_{i=0}^{n_Z} \sigma_i(x) h^Z_i(x): \sigma_i \in \Sigma_{r-r^Z_i}[x], i=0,\ldots,n_Z \big\}.
\end{align*}



A set $\Omega = \{x\in \R^n: h_i(x) \geq 0,h_i(x) \in \R[x], i=1,\ldots,n_{\Omega} \} \subseteq \R^n$  is said to satisfy Putinar's condition if there exists $\sigma \in \R[x]$ such that $\sigma = \sigma_0 + \sum_{i=1}^{n_{\Omega}} \sigma_i h_i$ for some $\{\sigma_i\}_{i=0}^{n_{\Omega}} \subset \Sigma [x]$, and the level set $\{x\in \R^n: \sigma(x) \geq 0\}$ is compact.
Putinar's condition can be satisfied by including the polynomial $N - ||x||_2^2$, where $N$ is a sufficiently large real number, in the defining polynomials $\{h_i\}$.
Here $x$ represents a general variable with $n$ components, so for the set $X\times U$, we consider the field $\R^{m+n}$, the polynomial ring $\R[x,u]$, and the cone $\Sigma[x,u]$.

\section{Optimization formulation}
\subsection{Discrete-time controlled Liouville equation}

The Liouville equation for continuous-time systems is a partial differential equation describing the evolution of the system state over time.
The discrete-time analogue of the Liouville equation was studied in Markov decision process, and was incorporated into the occupation measure approach in \cite{korda2014convex,magron2017semidefinite,savorgnan2009discrete}.
For our controller synthesis purpose, we are going to propose a new form of the Liouville equation, which we call the \textit{discrete-time controlled Liouville equation}.

Given measurable spaces $(X_1,\mc{A}_1)$ and $(X_2,\mc{A}_2)$, a measurable function $p: X_1\to X_2$ and a measure $\nu: \mc{A}_1\to [0,+\infty]$, the pushforward measure of $\nu$ is defined to be $p_*\nu: \mc{A}_2\to [0,+\infty]$
\[
p_*\nu(A):=\nu(p^{-1}(A))
\]
for all $A \in \mc{A}_2$.
Define $\pi$ to be the projection map from $X\times U$ to $X$, i.e., $\pi: X\times U \to X, (x,u) \mapsto x$.
The system dynamics $\phi: X\times U \to X$ is as defined in the previous section.
Let $X_0,X_T \subseteq X$ be the measurable sets containing all possible initial states and final states of the system, respectively.
The discrete-time controlled Liouville equation is 
\begin{align}
        \mu + \pi_* \nu = \phi_* \nu + \mu_0, \label{Liouville}
\end{align}
where $\mu_0 \in \mc{M}_+(X_0),\mu\in \mathcal{M}_+(X_T)$ and $\nu \in \mathcal{M}_+(X\times U)$.

We can view the initial measure $\mu_0$ as the distribution of the mass of the initial states of the system trajectories (not necessarily normalized to 1), the occupation measure $\nu$ as describing the volume occupied by the trajectories, and the final measure $\mu$ as the distribution of the mass of the final states of the system trajectories. 
For example, $\mu_0 = \delta_{x_0}$, $\nu = \delta_{(x_0,u_0)} + \ldots + \delta_{(x_{T-1},u_{T-1})}$, and $\mu = \delta_{x_T}$ is a solution to the controlled Liouville equation, describing the system trajectory $\{x_0,x_1 = \phi(x_0,u_0),\ldots,x_T = \phi(x_{T-1},u_{T-1})\}$, where $\delta_x$ is the Dirac measure centered at $x$.
It is possible that the measure $\nu$ can be disintegrated as $\nu_1(du|x)\nu_2(dx)$ for some measure $\nu_2$ on $X$ and some probability measure $\nu_1(du|x)$ on $U(x)$ for every $x \in X$, as noted in \cite{savorgnan2009discrete}.

\subsection{Primal-dual infinite-dimensional LP}

We formulate the infinite-dimensional LP on measures as follows:
\begin{align}
    \begin{aligned}
    \sup &\  \int_{X} 1 d\mu_0 \\
        \text{s.t. } &\  \mu + \pi_* \nu = \phi_* \nu + \mu_0, \\
        &\ \mu_0 + \hat{\mu}_0 = \lambda_{X},\\
        &\ \mu_0,\hat{\mu}_0 \in \mathcal{M}_+(X),\mu \in  \mathcal{M}_+(Z),\\
        &\ \nu \in \mathcal{M}_+(X\times U).
    \end{aligned}\label{lp_primal}
\end{align}

The objective is to maximize the mass of the initial measure.
The first constraint is the controlled Liouville equation.
Notice that we require the final measure $\mu$ to be supported on $Z$. This constraint, together with the objective, means that we want as many system trajectories as possible to land in $Z$.
The second constraint ensures that the initial measure is dominated by the Lebesgue measure on $X$, and if the optimal solution is achieved, then the initial measure would be the Lebesgue measure on a set of initial states whose trajectories end up in $Z$ and the optimal value is the volume of the set (similar to the idea in Theorem 3.1 in \cite{henrion2009approximate}).

The dual LP on continuous functions is given by 

\begin{align}
    \begin{aligned}
        \inf & \ \int_X w(x) d\lambda_X \\
        \text{s.t.} & \ v(x) - v(\phi(x,u)) \geq 0, \forall x\in X, \forall u\in U, \\
        & \ w(x) - v(x) - 1 \geq 0, \forall x \in X, \\
        & \ w(x) \geq 0, \forall x \in X, \\
        & \ v(x) \geq 0, \forall x \in Z, \\
        & \ v,w \in \mathcal{C}(X). 
    \end{aligned}\label{lp_dual}
\end{align}

\section{Semidefinite relaxations}
We have formulated the infinite-dimensional LP on measures and its dual on continuous functions, but we cannot solve them directly.
A practical solution is to approximate the original LP by a family of finite-dimensional SDP's.
This relaxation is based on the idea that measures can be characterized by their moments, just as signals can be characterized by their Fourier coefficients.
By solving the relaxed SDP's of certain degrees, we can extract controllers in the form of polynomials of corresponding degrees.
In this section, we first introduce some background knowledge on moments of measures. 
For more detailed treatments, please refer to \cite{lasserre2010moments}.
Next we formulate the relaxed SDP's on moments of measures and their dual on SOS polynomials.
Finally, we show how to extract controllers from the SDP solutions.

\subsection{Preliminaries} 

Any polynomial $p(x)\in \R[x]$ can be expressed in the monomial basis as
\[
    p(x) = \sum_{\alpha} p_\alpha x^{\alpha},
\]
where $\alpha \in \N^n$, and 
$p(x)$ can be identified with its vector of coefficients $p:=(p_\alpha)$ indexed by $\alpha$.
Any measure $\mu$ is characterized by its sequence of moments, defined by
\[
\int x^\alpha d\mu, \alpha \in \N^n. 
\]
Given a sequence of real numbers $y:=(y_\alpha)$, we define the linear functional $\ell_y: \R[x] \to \R$ by 
\[
    \ell_y(p(x)):= p^\top y = \sum_\alpha p_\alpha y_\alpha.
\]
If $y = (y_\alpha)$ is a sequence of moments for some measure $\mu$, i.e.,
\[
    y_\alpha = \int x^\alpha d\mu,
\]
then $\mu$ is called a representing measure for $y$.
If $y$ has a representing measure $\mu$, then the linear function $\ell_y$ is the same as integration with respect to $\mu$:
\[
\int p d\mu = \int \sum_\alpha p_\alpha x^\alpha d\mu = \sum_\alpha p_\alpha \int x^{\alpha} d\mu = \ell_y(p(x)).
\]

Given $r\in \N$, define $\N^n_r = \{\beta\in \N^n: |\beta|:=\sum_i \beta_i \leq r\}$.
Define the moment matrix $M_r(y)$ of order $r$ with entries indexed by multi-indices $\alpha$ (rows) and $\beta$ (columns)
\[
    [M_r(y)]_{\alpha,\beta}:= \ell_y(x^\alpha x^\beta) = y_{\alpha+\beta}, \forall \alpha,\beta \in \N^n_r.
\]
If $y$ has a representing measure, then $M_r(y) \succeq 0$, $\forall r\in \N$.
However, the converse is generally not true.

Given a polynomial $u(x)\in \R[x]$ with coefficient vector $u = (u_\gamma)$, define the localizing matrix w.r.t. $y$ and $u$ to be the matrix indexed by multi-indices $\alpha$ (rows) and $\beta$ (columns)
\begin{align*}
    [M_r(uy)]_{\alpha,\beta} &:= \ell_y(u(x)x^\alpha x^\beta) \\
    &= \sum_{\gamma} u_{\gamma}y_{\gamma + \alpha + \beta}, \forall \alpha,\beta \in \N^n_r.
\end{align*}
If $y$ has a representing measure $\mu$, then $M_r(uy)\succeq 0$ whenever the support of $\mu$ is contained in $\{x\in \R^n: u(x) \geq 0\}$.
Conversely, if $X$ is a compact semi-algebraic set as defined in Section II, if $X$ satisfies Putinar's condition, and if $M_r(h^X_j y) \succeq 0, j=0,\ldots,n_X, \forall r$, then $y$ has a finite Borel representing measure with support contained in $X$ (Theorem 3.8(b) in \cite{lasserre2010moments}).

\subsection{Primal-dual finite-dimensional SDP}
For each $r \geq r_{min} := \max_{i, j, k} \{r^X_i, r^U_j, r^Z_k \}$, let $y_0 = (y_{0,\beta}),\beta\in \N^n_{2r}$, be the finite sequence of moments up to degree $2r$ of the measure $\mu_0$.
Similarly, $y_1, \hat{y}_0, y^{X}$, and $z$ are finite sequences of moments up to degree $2r$ associated with measures $\mu,\hat{\mu}_0$, $\lambda_X$, and $\nu$, respectively.
Let $d := $ degree  $\phi$.
The infinite-dimensional LP on measures (\ref{lp_primal}) can be relaxed with the following semidefinite program on moments of measures:

\begin{align}
    \begin{aligned}
        \sup   &\  y_{0,0}  \\
        \text{s.t.} & \ y_{1,\beta} + \ell_z(x^\beta) = \ell_z(\phi(x,u)^\beta) + y_{0,\beta}, \forall \beta \in \N^n_{2r}, \\
        & \ y_{0,\beta} + \hat{y}_{0,\beta} = y^{X}_\beta, \forall \beta \in \N^n_{2r},\\
        & \ \mathbf{M}_{r-r^X_j}(h^X_j y_0) \succeq 0, j = 1,\ldots,n_X,\\
        & \ \mathbf{M}_{r-r^X_j}(h^X_j \hat{y}_0) \succeq 0,j = 1,\ldots,n_X,\\
        & \ \mathbf{M}_{rd-r^X_j}(h^X_j z) \succeq 0,j = 1,\ldots,n_X,\\
        & \ \mathbf{M}_{rd-r^U_j}(h^U_j z) \succeq 0,j = 1,\ldots,n_U,\\
        & \ \mathbf{M}_{r-r^Z_j}(h^Z_j y_1) \succeq 0,j = 1,\ldots,n_Z. 
    \end{aligned}\label{sdp_primal}
\end{align}

The dual of (\ref{sdp_primal}) is the following SDP on polynomials of degrees up to $2r$:

\begin{align}
    \begin{aligned}
        \inf_{v,w} & \ \sum_{\beta\in\N^n_{2r}} w_\beta y_\beta^X  \\
        \text{s.t.} & \ v - v\circ \phi   \in \mathbf{Q}^{XU}_{rd}, \\
        & \ w - v - 1\in \mathbf{Q}^X_{r}, \\
        & \ w \in \mathbf{Q}^X_{r}, v \in \mathbf{Q}^Z_{r}, \\
        & \  v,w \in \R_{2r}[x],   
    \end{aligned}\label{sdp_dual}
\end{align}
where $\circ$ denotes function composition.
The dual SDP (\ref{sdp_dual}) is a strengthening of the dual LP (\ref{lp_dual}) by requiring nonnegative polynomials in (\ref{lp_dual}) to be SOS polynomials up to certain degrees.

\subsection{Controller extraction}
The controllers can be extracted from the primal SDP (\ref{sdp_primal}) as in \cite{majumdar2014convex,korda2014controller}.
We describe the procedure in detail in the following.

Fix $r\in \N$ in the SDP's (\ref{sdp_primal}) and (\ref{sdp_dual}).
Let each $u_i$ be a degree-$r$ polynomial in $x$, $i = 1,\ldots,m$.
Identify $u_i$ with its vector of coefficients $(u_{i,\alpha})$. 
$\nu$ is a measure supported on $X\times U$.
By solving the primal SDP (\ref{sdp_primal}), we obtain the moments of $\nu$ (as subsequences of $z$):
\begin{align*}
    &\tau_{i,\alpha} := \int x^\alpha u_i d\nu, \forall \alpha \in \N_r^n,\\
    &\rho_{\alpha} := \int x^\alpha d\nu, \forall \alpha \in \N_r^n.
\end{align*}
Then 
\[
    M_r(\rho) \cdot (u_{i,\alpha})_{\alpha} = (\tau_{i,\alpha})_{\alpha},
\]
where $(u_{i,\alpha})_{\alpha}$ is the column vector of coefficients of the polynomial $u_i(x)$ indexed by $\alpha$, and $(\tau_{i,\alpha})_{\alpha}$ is the column vector consisting of $\tau_{i,\alpha}$'s indexed by $\alpha$.
The controller $u_i(x)$ can be approximated by taking the pseudo-inverse of the moment matrix $M_r(\rho)$:
\[
    (u_{i,\alpha})_{\alpha} =[M_r(\rho)]^+\cdot  (\tau_{i,\alpha})_{\alpha}.
\]

As noted in \cite{korda2014controller}, the approximated controller does not always satisfy the control input constraints.
The easiest remedy is to limit the control input to be the boundary values, $\pm 1$, if the constraints are violated.
For all the examples in the Examples Section, we used this method.
Most of the time, the control input constraints were not violated.
Another method is to solve an SOS optimization problem as in \cite{korda2014controller}.

In general, our controller synthesis method is heuristic.
The controllable region needs to be checked a posteriori.
In the next section, we show that we can over-approximate the controllable region using a simplified form of our optimization formulation.

\section{Uncontrolled case: outer approximation of the backward reachable set}
In this section, we consider a special case -- the discrete-time autonomous polynomial system
\[ x_{t+1} = f(x_t), \]
where $X$, $f(x)$, and the target set $Z$ are defined as before.
Given a time step $T \in \N$,
define the $T$-step backward reachable set 
\begin{align*}
    X_0^T := \{ & x_0\in X: x_t = f(x_{t-1}) = \cdots = f^{t}(x_0) \in Z\\ &\text{ for some } 0 \leq t \leq T,
     \text{and } x_{t_i} \in X, \forall 0 \leq t_i \leq t\}.
\end{align*}
This is the set of points in $X$ that enter the target region $Z$ within $T$ time steps and whose trajectories do not leave $X$ before entering $Z$.
Once a point enters $Z$, what happens to it next is not our concern.
We are going to over-approximate the backward reachable set
\[
    X_0^{\infty} := \bigcup_{T=0}^{\infty} X_0^T,
\]
which is the union of all points in $X$ that enter $Z$ in finite time.
Denote by $\bar{X}_0^\infty$ the closure of $X_0^{\infty}$.

The primal LP is obtained from LP (\ref{lp_primal}) by modifying the Liouville equation to be the same as the one in \cite{magron2017semidefinite} and modifying the support of the occupation measure $\nu$ to be $X$.
The primal and dual LP's are formulated as follows
\begin{align}
    \begin{aligned}
    p := \sup &\  \int_{X} 1 d\mu_0 \\
        \text{s.t. } &\  \nu + \mu = f_{*} \nu + \mu_0, \\
        &\ \mu_0 + \hat{\mu}_0 = \lambda_{X},\\
        &\ \mu_0,\hat{\mu}_0,\nu \in \mathcal{M}_+(X),\\
        &\ \mu \in  \mathcal{M}_+(Z).
    \end{aligned}\label{special_lp_primal}
\end{align}
\begin{align}
    \begin{aligned}
        d := \inf & \ \int_X w(x) d\lambda_X \\
        \text{s.t.} & \ v(x) - v(f(x)) \geq 0, \forall x \in X, \\
        & \ w(x) - v(x) -1 \geq 0, \forall x \in X, \\
        & \ w(x) \geq 0, \forall x \in X, \\
        & \ v(x) \geq 0, \forall x \in Z, \\
        & \ v,w \in \mathcal{C}(X). 
    \end{aligned}\label{special_lp_dual}
\end{align}
\begin{proposition}{\ptcnt{cnt}}
Suppose there exists a constant $M > 0$ such that for any feasible solution $(\mu_0,\hat{\mu}_0,\nu,\mu)$ of the LP (\ref{special_lp_primal}), the mass of $\nu$ is bounded by $M$, i.e., $\int_X 1 d\nu < M$.

(a) If $f(X) \subseteq X$, then LP (\ref{special_lp_primal}) admits an optimal solution $(\mu_0^*,\hat{\mu}_0^*,\nu^*,\mu^*)$ such that $\mu_0^* = \lambda_{{X}_0^\infty}$ and $p^* = \vol {X}_0^\infty$.

(b) There is no duality gap between the primal LP (\ref{special_lp_primal}) and the dual LP (\ref{special_lp_dual}). \hfill \qedsymbol
\end{proposition}

The semidefinite relaxations can be obtained similarly.
The primal is
\begin{align}
    \begin{aligned}
        p_r &:= \sup_{\substack{y_0,\hat{y}_0,z,a}}   y_{0,0} \\
        \text{s.t.} & \  y_{1,\beta} + z_\beta  = \ell_z(f(x)^\beta) + y_{0,\beta}, \forall \beta \in \N^n_{2r}, \\
        & \ y_{0,\beta} + \hat{y}_{0,\beta} = y^{X}_\beta, \forall \beta \in \N^n_{2r},\\
        & \ \mathbf{M}_{r-r^X_j}(h^X_j y_0) \succeq 0, j = 1,\ldots,n_X,\\
        & \ \mathbf{M}_{r-r^X_j}(h^X_j \hat{y}_0) \succeq 0,j = 1,\ldots,n_X,\\
        & \ \mathbf{M}_{rd-r^X_j}(h^X_j z) \succeq 0,j = 1,\ldots,n_X,\\
        & \ \mathbf{M}_{r-r^Z_j}(h^Z_j y_1) \succeq 0,j = 1,\ldots,n_Z.\\
    \end{aligned}\label{special_sdp_primal}
\end{align}
The dual is 
\begin{align}
    \begin{aligned}
        d_r := \inf_{v,w} & \ \sum_{\beta\in\N^n_{2r}} w_\beta y_\beta^X \\
        \text{s.t.} & \ v - v\circ f \in \mathbf{Q}^X_{rd}, \\
        & \ w - v - 1\in \mathbf{Q}^X_{r}, \\
        & \ w \in \mathbf{Q}^X_{r}, v \in \mathbf{Q}^Z_{r}, \\
        & \ v,w \in \R_{2r}[x].         
    \end{aligned}\label{special_sdp_dual}
\end{align}

\begin{proposition}{\ptcnt{cnt}}
Let $r \geq r_{min}$.

(a) 
The primal SDP (\ref{special_sdp_primal}) and the dual SDP (\ref{special_sdp_dual}) are both feasible.
If the primal SDP (\ref{special_sdp_primal}) has a strictly feasible solution, then there is no duality gap between the primal SDP (\ref{special_sdp_primal}) and the dual SDP (\ref{special_sdp_dual}), and the optimal value of SDP (\ref{special_sdp_dual}) is attained.

(b) 
Let $(v_r,w_r)$ be a feasible solution to SDP (\ref{special_sdp_dual}).
Define
\[
X_{0r} = \{x\in X |w_r(x) -1 \geq 0\}.
\]
Then $X_{0r} \supseteq \bar{X}^\infty_0 \supseteq {X}^\infty_0$.
Suppose the conditions in Part (a) hold.
In addition, if there exists a sequence of polynomials $(u_k)_{k\in\N}$ satisfying (i) $u_k > 1_{\bar{X}^\infty_0}$ on $X$, (ii) $(u_k)$ converges to $1_{\bar{X}^\infty_0}$ in $L^1$ norm, and (iii) $u_k(x) - u_k(f(x)) > 0, \forall x\in X$, then SDP (\ref{special_sdp_dual}) has an optimal solution $(v_r,w_r)$ such that 
\[
\lim_{r\to\infty} \int_X |w_r(x) - 1_{\bar{X}_0^\infty}(x)|d\lambda_X = 0.
\]
\hfill\qedsymbol
\end{proposition}

\noindent \textit{Remark.}
Part (a) is a standard strong duality theorem for SDP's.
Part (b) indicates that $X_{0r}$ is an outer approximation of the closure of the backward reachable set.
If we define
\[
\widetilde{X}_{0r} = \bigcap_{k=r_{min}}^r X_{0k},
\]
then the approximation by the sequence of sets $\{\widetilde{X}_{0r}\}_r$ is monotone.
The last technical condition in Part (b) can be understood as follows.
Since $\bar{X}_0^\infty$ is closed, the indicator function $1_{\bar{X}_0^\infty}$ is upper semi-continuous.
So there exists a decreasing sequence of bounded continuous functions $(u_k)_{k\in \N}$ converging pointwise to $1_{\bar{X}_0^\infty}$ on $X$.
By the Dominated Convergence Theorem, $(u_k)_{k\in \N}$ converges to $1_{\bar{X}_0^\infty}$ in $L^1$ norm.
By the Stone-Weierstrass Theorem, each $u_k$ can be approximated uniformly arbitrarily well by polynomials.
Therefore, there exists a sequence of polynomials $(\tilde{u}_k)_{k\in \N}$ satisfying conditions (i) and (ii).
So (iii) is an additional constraint.
If (iii) holds, then Putinar's Positivstellensatz implies that SDP (\ref{special_sdp_dual}) has a feasible solution whose $w$-component resembles $(\tilde{u}_k)$.
This establishes the vanishing error of the hierarchical SDP approximations.
\hfill \qedsymbol

In practice, however, given a system it is not known a priori if condition (iii) holds or not.
Even if it is known, current numerical solvers can only handle SDP's up to a certain degree. 
Whether the approximation up to that degree is good or not is not known.

While the approach approximates the backward reachable set of autonomous systems, it can also approximate the backward controllable set of systems subject to polynomial state feedback control inputs.
This is immediately seen by plugging the polynomial control law $u_t = u(x_t)$ into the control affine polynomial system $x_{t+1} = f(x_t) + g(x_t)u_t = f(x_t) + g(x_t)u(x_t)$, yielding a polynomial closed-loop dynamical system.

\section{Examples}
We illustrate our methods on five discrete-time polynomial systems.
All computations are done using MATLAB 2016b, the SDP solver MOSEK 8, and the polynomial optimization toolbox Spotless \cite{tobenkin2013spotless}.

\subsection{Van der Pol oscillator}
In this example, we are going to over approximate the backward reachable set of the uncontrolled reversed-time Van der Pol oscillator (Example 9.2 in \cite{henrion2014convex}) given by 
\begin{align*}
    &\dot{x}_1 = -2x_2,\\
    &\dot{x}_2 = 0.8x_1 + 10(x_1^2 -0.21)x_2.
\end{align*}
Discretizing the model with the explicit Euler scheme with a sampling time $\delta t = 0.01$, the discrete-time system is 
\begin{align*}
    &x_1^+ = (-2x_2)\delta t + x_1,\\
    &x_2^+ = (0.8x_1 + 10(x_1^2 -0.21)x_2) \delta t + x_2.
\end{align*}
Choose $X = \{x \in \R^2: |x_1|^2 \leq 1.5^2, |x_2|^2 \leq 1.5^2\}$ and $Z = \{x\in \R^2:|x_1|^2 \leq 0.1^2, |x_2|^2\leq 0.1^2\}$.

\begin{figure}[H]
    \begin{tabular}{@{}c@{}}
  \includegraphics[width=.5\linewidth]{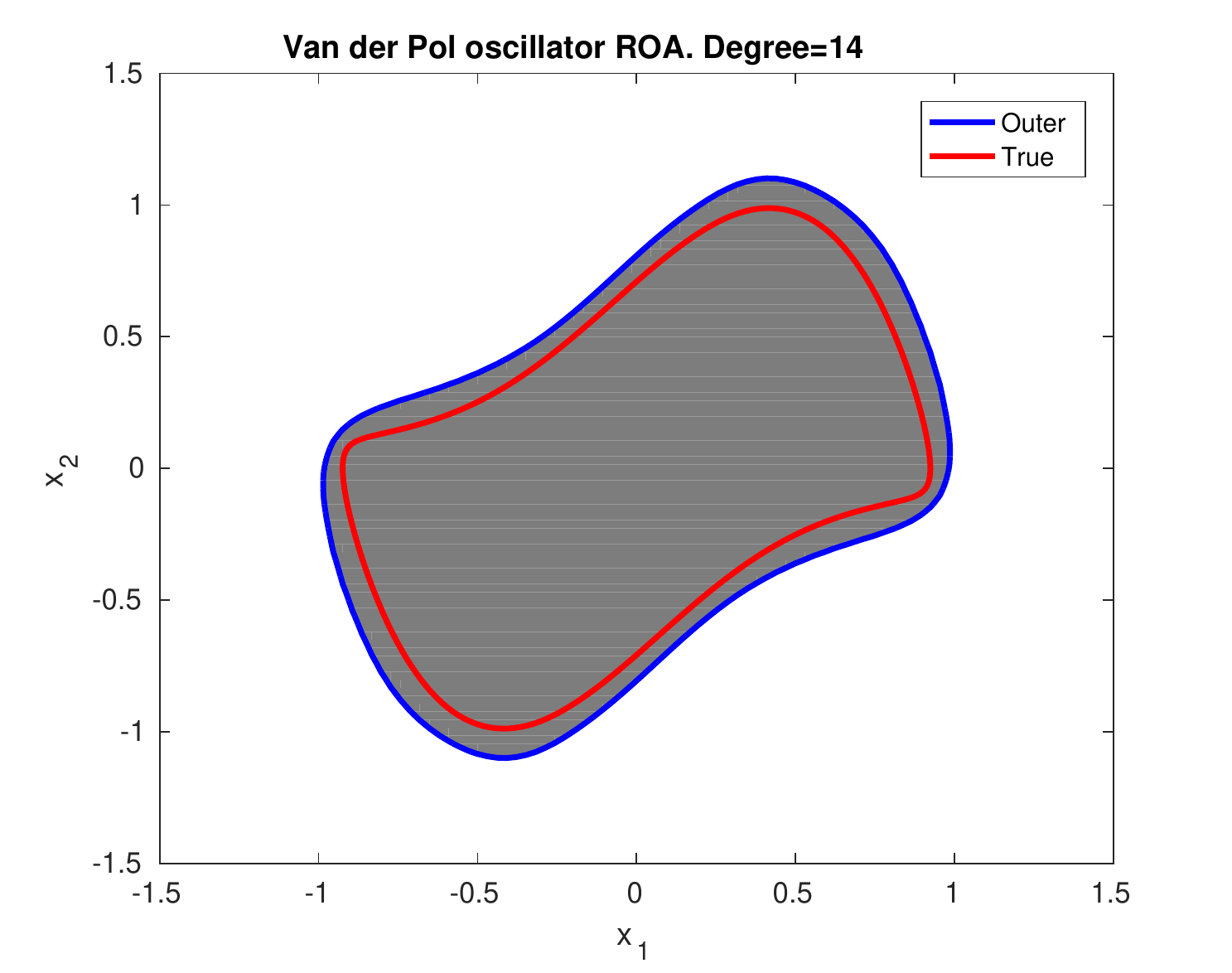}
  \includegraphics[width=.5\linewidth]{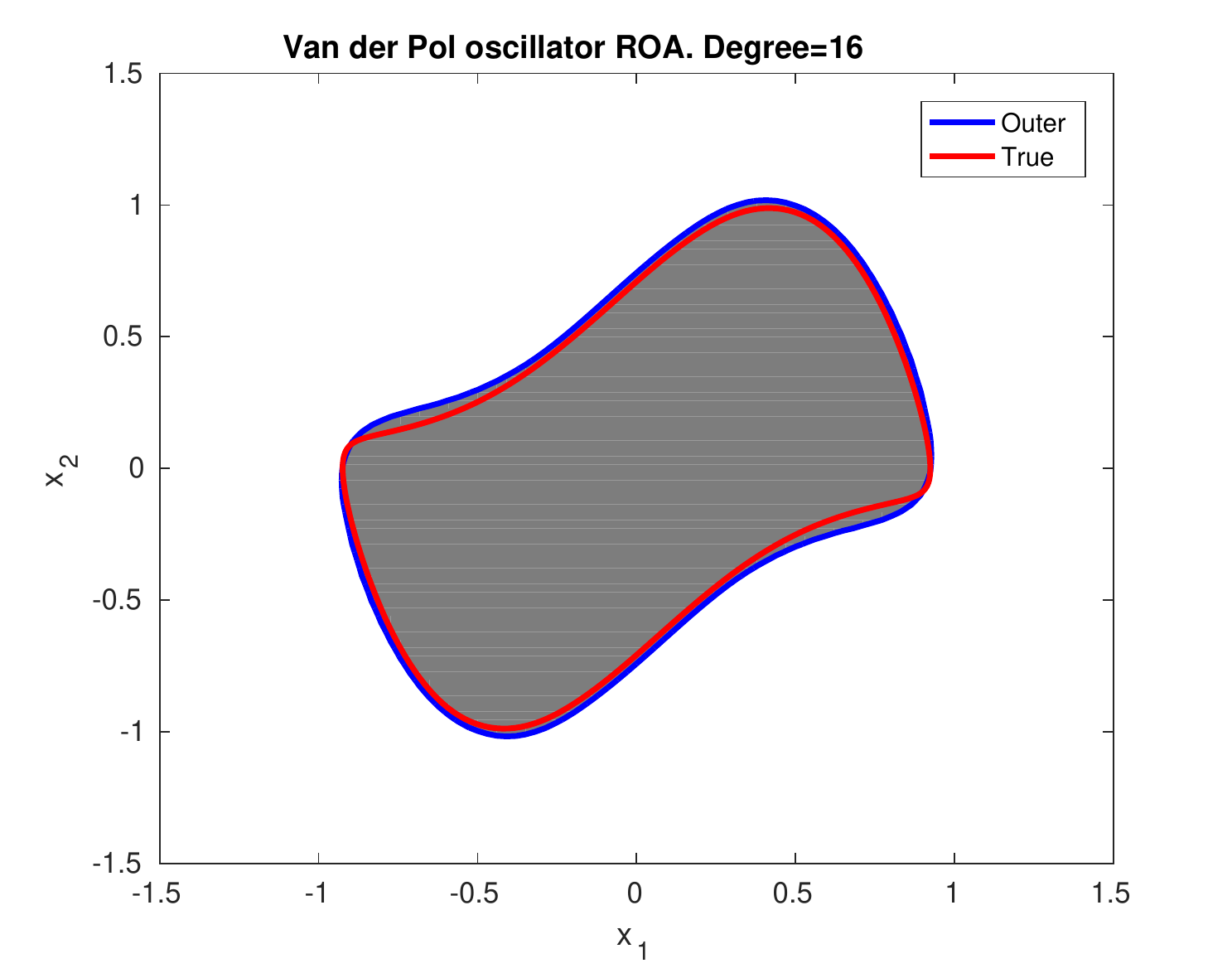}
    \end{tabular}
  \caption{Degree-14 and degree-16 outer approximations to the backward reachable set (or region of attraction).}
  \label{fig:vanderpol}
\end{figure}

We approximate the backward reachable set by degree-14 and 16 polynomials.
As show in Figure \ref{fig:vanderpol}, the gray areas are the approximate backward reachable sets.
The areas enclosed by the red lines are the true backward reachable set, which was obtained analytically by integrating backwards in time.

\subsection{Double integrator}
Consider a double integrator discretized by the explicit Euler scheme with a sampling time $\delta t = 0.01$.
The discrete-time dynamics equations are 
\begin{align*}
    &x_1^+ = x_1 + 0.01x_2,\\
    &x_2^+ = x_2 + 0.01u.
\end{align*}
We are going to design controllers and then approximate the backward reachable set of the closed loop system.
We consider the state constraint set $X = \{x\in \R^2: |x_1|\leq 1, |x_2|\leq 1\}$, and the target set $Z = \{x\in \R^2: ||x||_2^2 \leq 0.05^2\}$.
We search for a degree-1 controller.

\begin{figure}[H]
  \includegraphics[width=.45\linewidth]{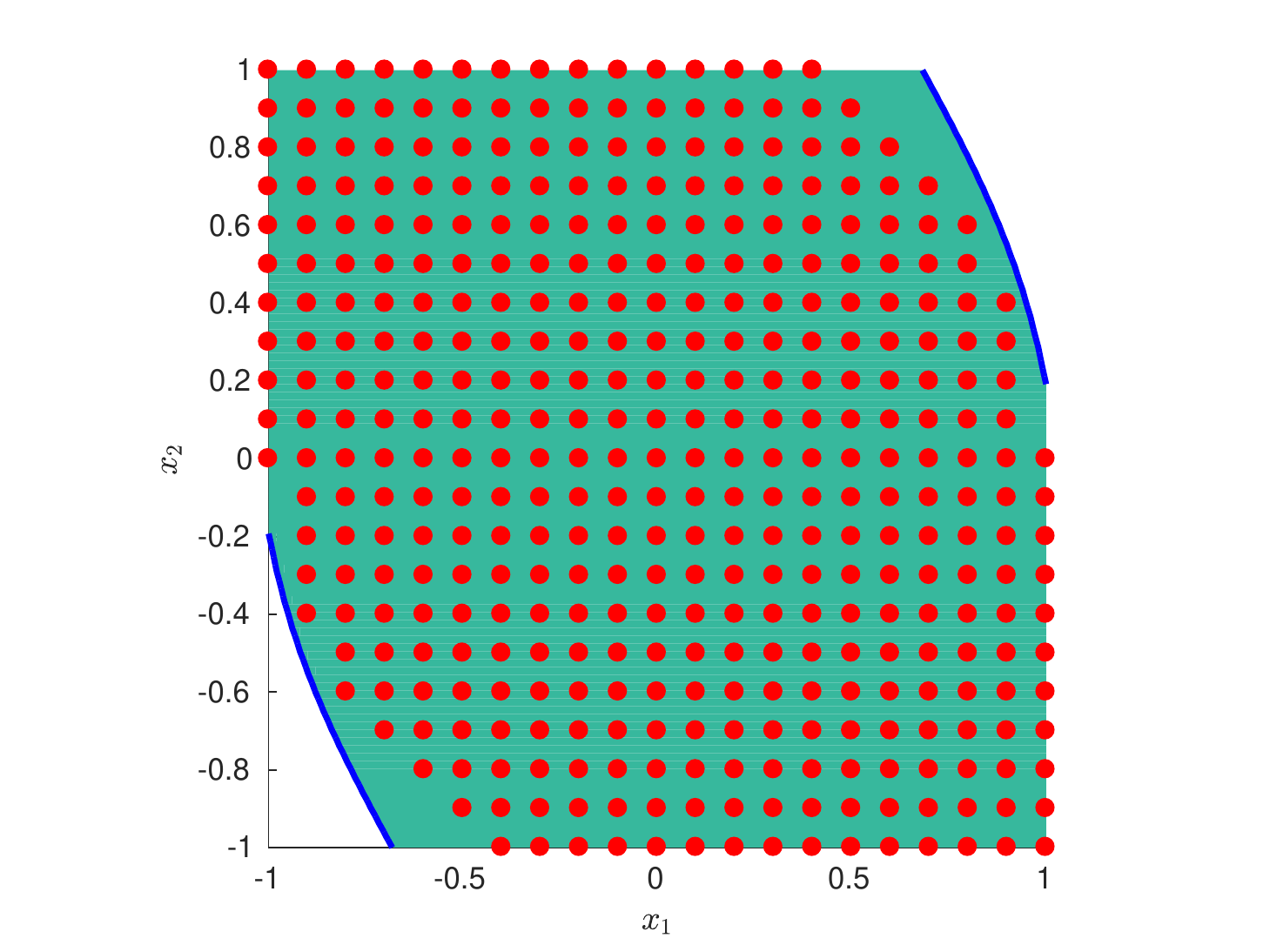}
  \includegraphics[width=.45\linewidth]{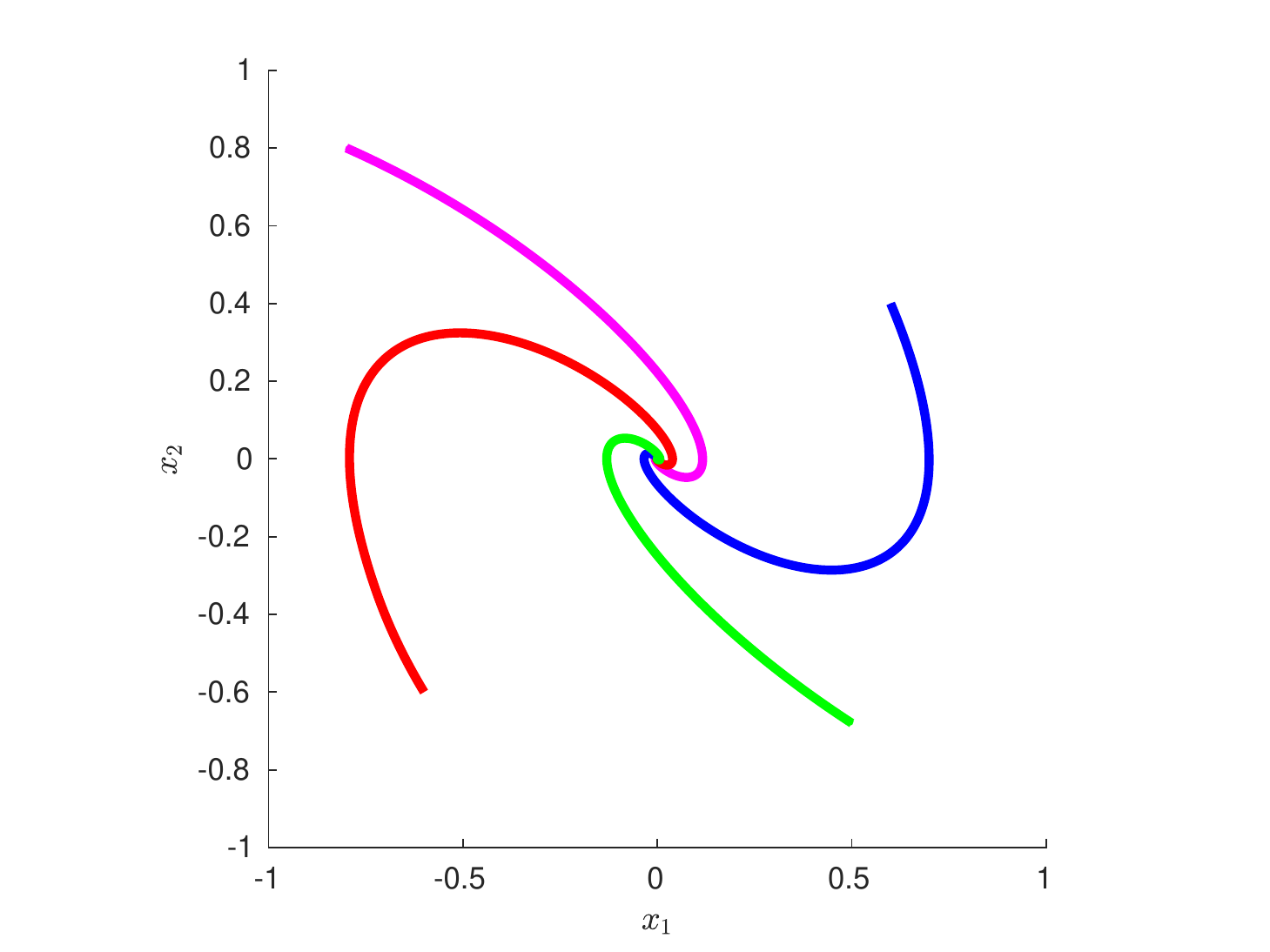}
  \caption{Left: The green area is a degree-10 approximation of the backward reachable set of the closed loop system. The red markers represent the controllable sample points. Right: Trajectories of four initial states under the extracted controller.}
  \label{fig:double_integrator}
\end{figure}

As shown in the left plot of Figure \ref{fig:double_integrator}, the green area is a degree-10 approximation of the backward reachable set of the closed loop system.
We cover $X$ by a uniform $20\times 20$ grid, and compute the trajectories of the grid vertices under the extracted controller.
The red markers represent the vertices that can be steered to $Z$ under the extracted controller in $T = 10^4$ time steps without violating state or control input constraints.
In the right plot of Figure \ref{fig:double_integrator}, we plotted the trajectories of four initial states, $(-0.8,0.8), (-0.6,-0.6), (0.6,0.4)$, and $(0.5,-0.68)$, under the extracted controller.

\subsection{Dubin's car}
Consider the Dubin's car model (Example 2 in \cite{majumdar2014convex})
\[
    \dot{a} = v\cos(\theta), \dot{b} = v\sin(\theta), \dot{\theta} = \omega,
\]
or by a change of coordinates, the Brockett integrator 
\[
    \dot{x}_1 = u_1, \dot{x}_2 = u_2, \dot{x}_3 = x_2 u_1 - x_1 u_2.
\]
The system has an uncontrollable linearization and does not admit any continuous time-invariant control law that makes the origin asymptotically stable \cite{devon2007kinematic}.
We are going to design a polynomial control law for the system.

\begin{figure}[H]
	\begin{minipage}{.35\linewidth}
  \includegraphics[width=\linewidth]{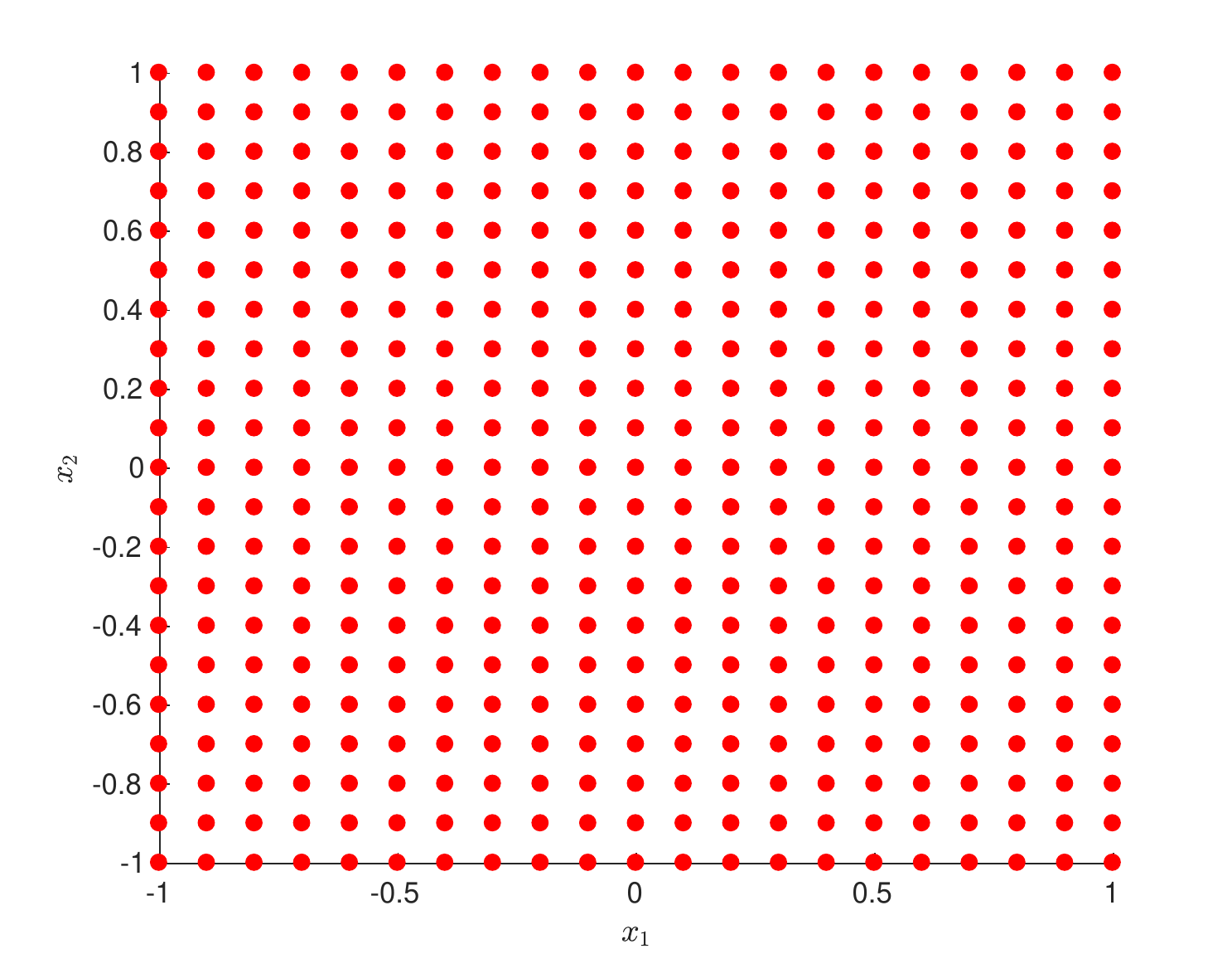}
  \includegraphics[width=\linewidth]{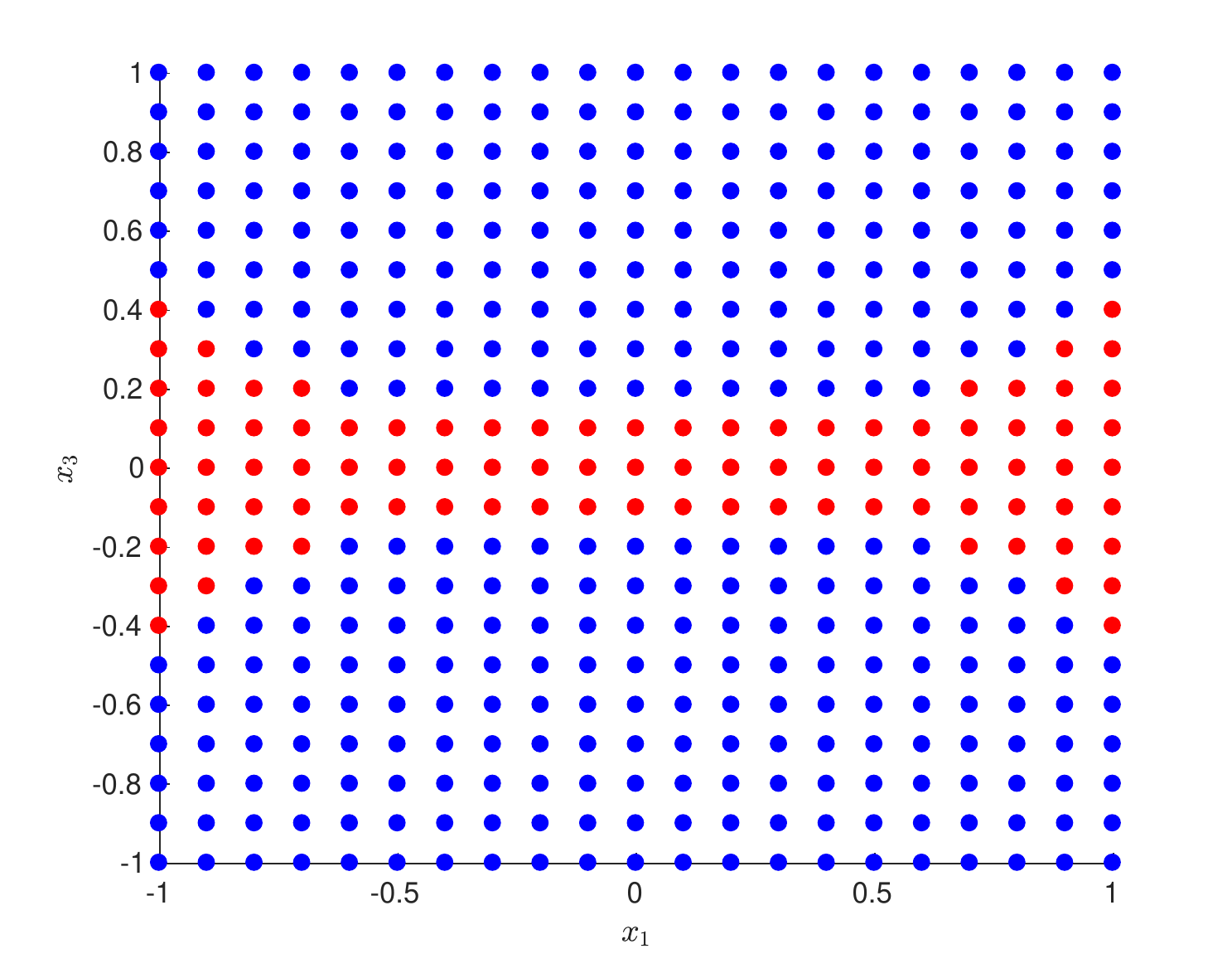}
	\end{minipage}
	\begin{minipage}{.63\linewidth}
  \includegraphics[width=\linewidth]{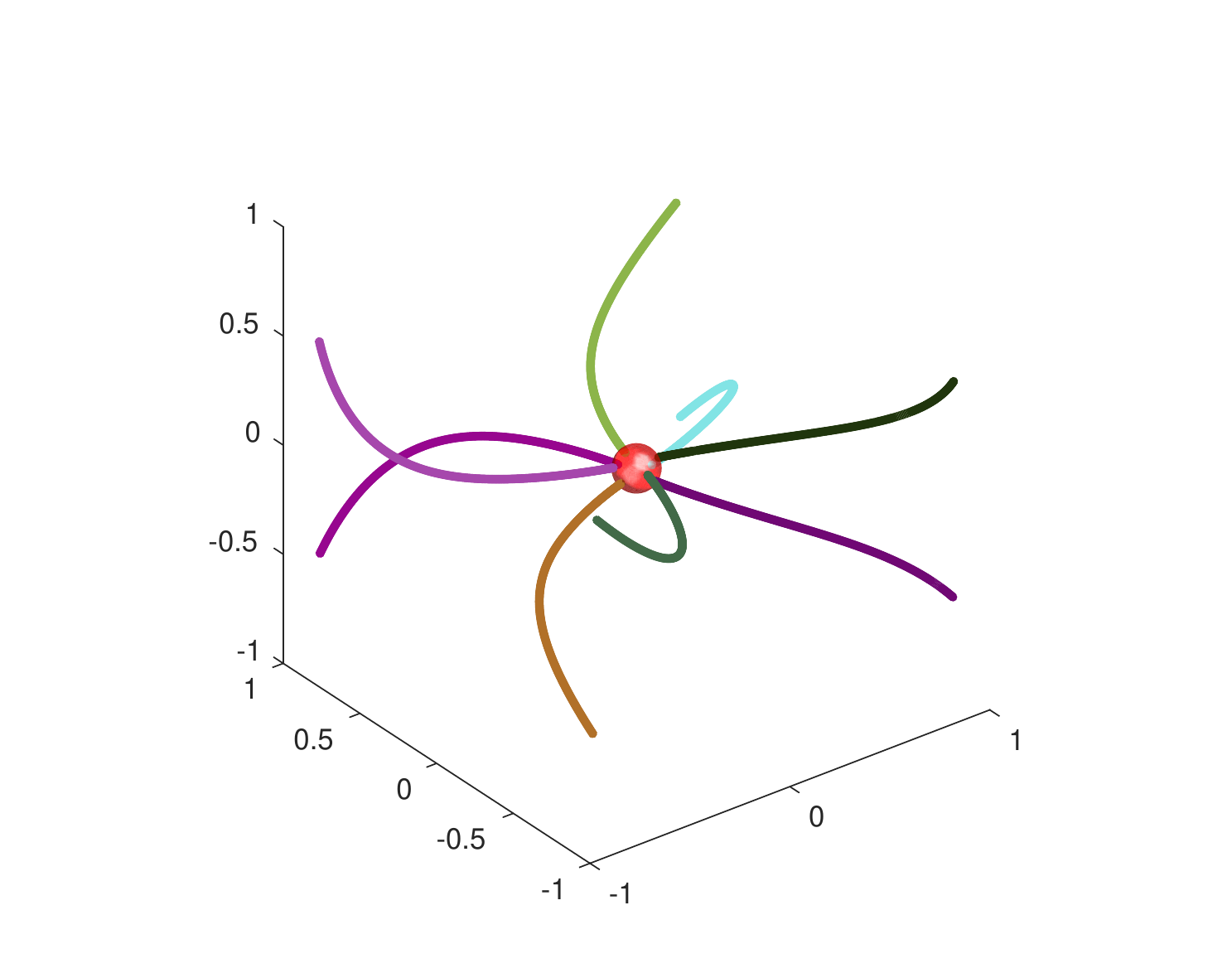}
	\end{minipage}
  \caption{Top left: The 2D section $\{x\in X: x_3 = 0\}$. Bottom left: The 2D section $\{x\in X: x_2 = 0\}$. The red vertices represent the initial states that can be regulated to the target set under the extracted controller. Right: Trajectories of the eight initial states $(\pm 0.9, \pm 0.9, \pm 0.5)$ under the extracted controller. The red ball in the center is the target set $Z$.}
  \label{fig:dubins_car}
\end{figure}

Discretize the system using the explicit Euler scheme with a sampling time $\delta t = 0.01$.
Choose $X = \{x\in \R^3: ||x||_\infty\leq 1\}$, and $Z = \{x\in \R^3: ||x||_2^2 \leq 0.1^2\}$.
We search for a degree-4 controller.
We sample the 2D sections $\{x\in X: x_3 = 0\}$ and $\{x\in X: x_2 = 0\}$ uniformly, and compute whether the grid vertices can be steered to $Z$ under the extracted controller in $10^4$ time steps.
In the left two plots of Figure \ref{fig:dubins_car}, the red vertices represent the initial states that can be regulated to the target set under the extracted controller, while the blue vertices are the rest.
The right plot of Figure \ref{fig:dubins_car} shows the trajectories of the eight initial states $(\pm 0.9, \pm 0.9, \pm 0.5)$ under the extracted controller.
They all reach the target set $Z$, represented by a red ball.
Some other initial states that cannot reach the target set actually end up somewhere very close to the target set.
For example the initial state $(0.8,-0.6,0.7)$ ends up at $(0,0,0.1224)$.

\subsection{Controlled 3D Van der Pol oscillator}
Consider the controlled 3D Van der Pol oscillator (Example 2 in \cite{korda2014controller}) discretized by the explicit Euler scheme with a sampling time $\delta t = 0.01$.
The dynamics are given by 
\begin{align*}
    &x_1^+ = x_1 - 2x_2 \delta t\\
    &x_2^+ = x_2 + (0.8 x_1 - 2.1 x_2 + x_3 + 10 x_1^2 x_2) \delta t\\
    &x_3^+ = x_3 + (-x_3 + x_3^3 + 0.5 u) \delta t
\end{align*}
Let the state constraint set be the unit ball $X = \{x\in \R^3: ||x||_2^2 \leq 1\}$ and the target set be $Z = \{x\in \R^3: ||x||_2^2 \leq 0.1^2\}$.
We search for a degree-1 controller, i.e., an affine controller.
\begin{figure}[H]
  \includegraphics[width=.45\linewidth]{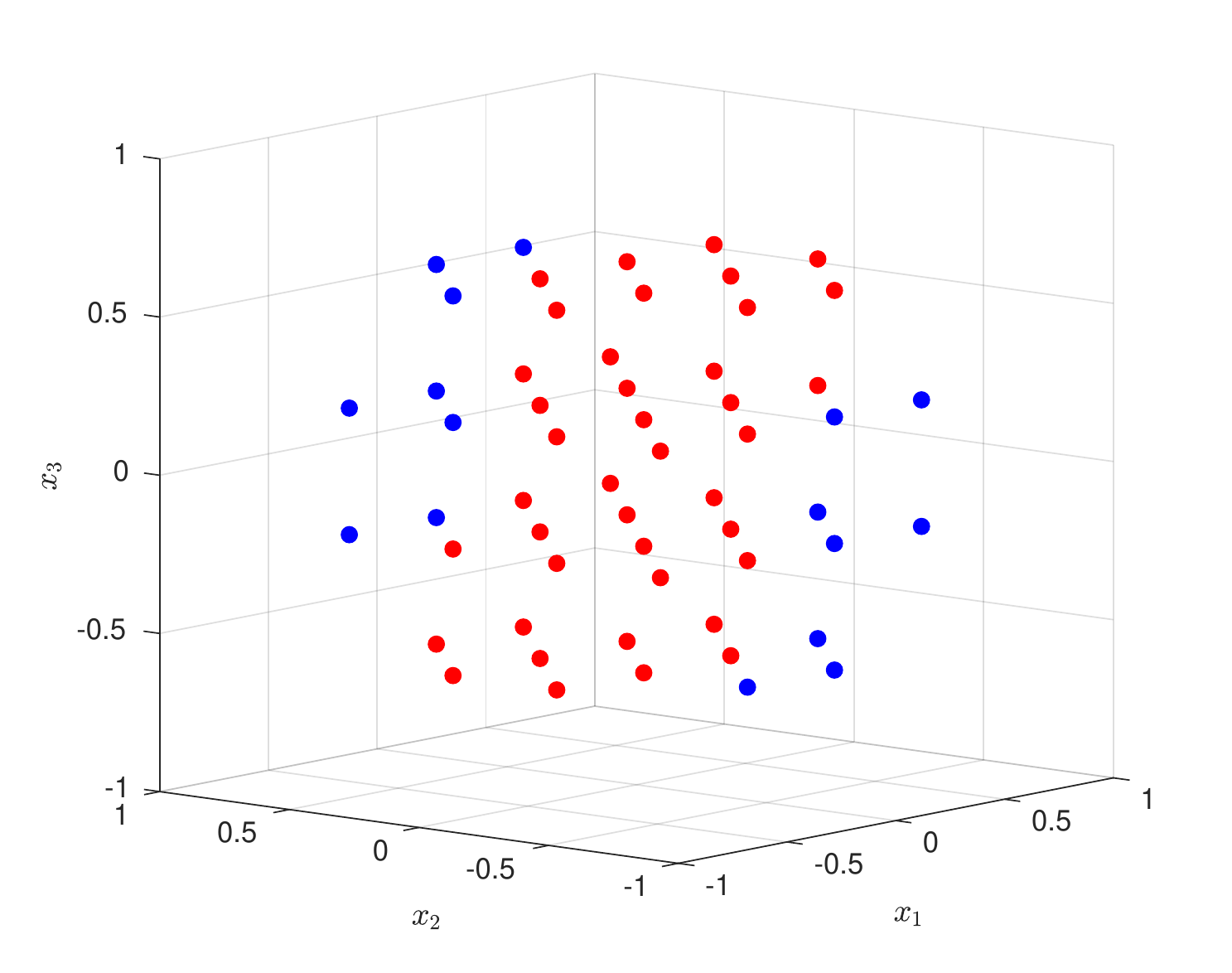}
  \includegraphics[width=.45\linewidth]{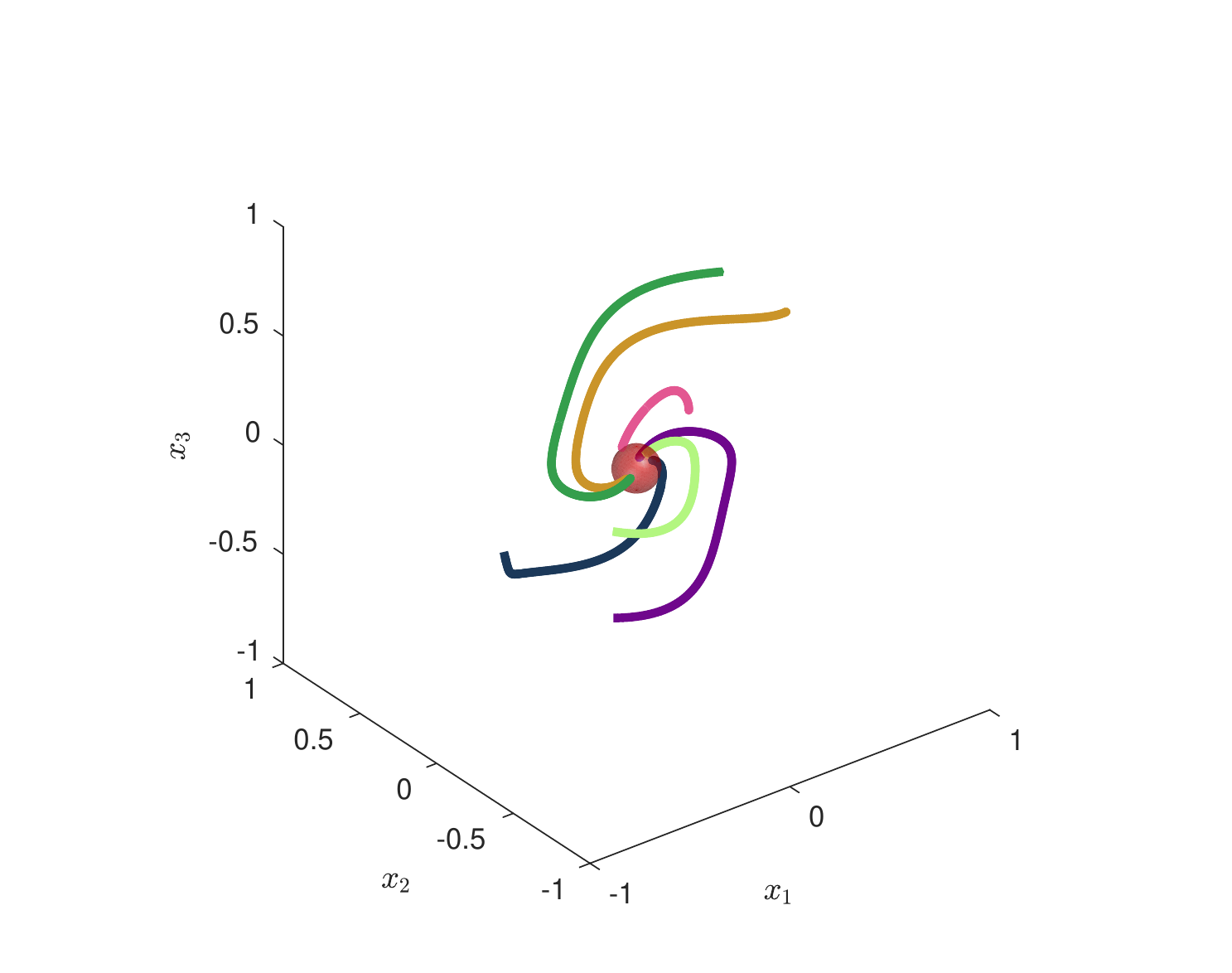}
  \caption{Left: Controllable sample points in red and uncontrollable points in blue. Right: Trajectories of six initial states under the extracted controller. The red ball in the center is the target set $Z$.}
  \label{fig:van_der_pol}
\end{figure}
We choose as our sample points the uniform $5 \times 5 \times 5$ grid vertices that are inside the unit ball $X$.
As shown in the left plot in figure \ref{fig:van_der_pol}, the red dots represent the sample points that can be controlled to the target set under the extracted controller in $10^4$ time steps.
The blue dots represent those cannot.
In the right plot, we show the trajectories of six initial states $(0.6,-0.6,-0.2)$, $(-0.6,-0.6,0.2)$,
$(0.6,0.2,0.6)$, $(0.6,-0.2,0.6)$, $(-0.2,0.6,-0.6)$, and $(-0.2,-0.6,0.6)$ under the extracted controller.
The red ball in the center represents the target set.

\subsection{Cart-pole system}
\begin{figure}[h]
    \centering
    \includegraphics[width=2in]{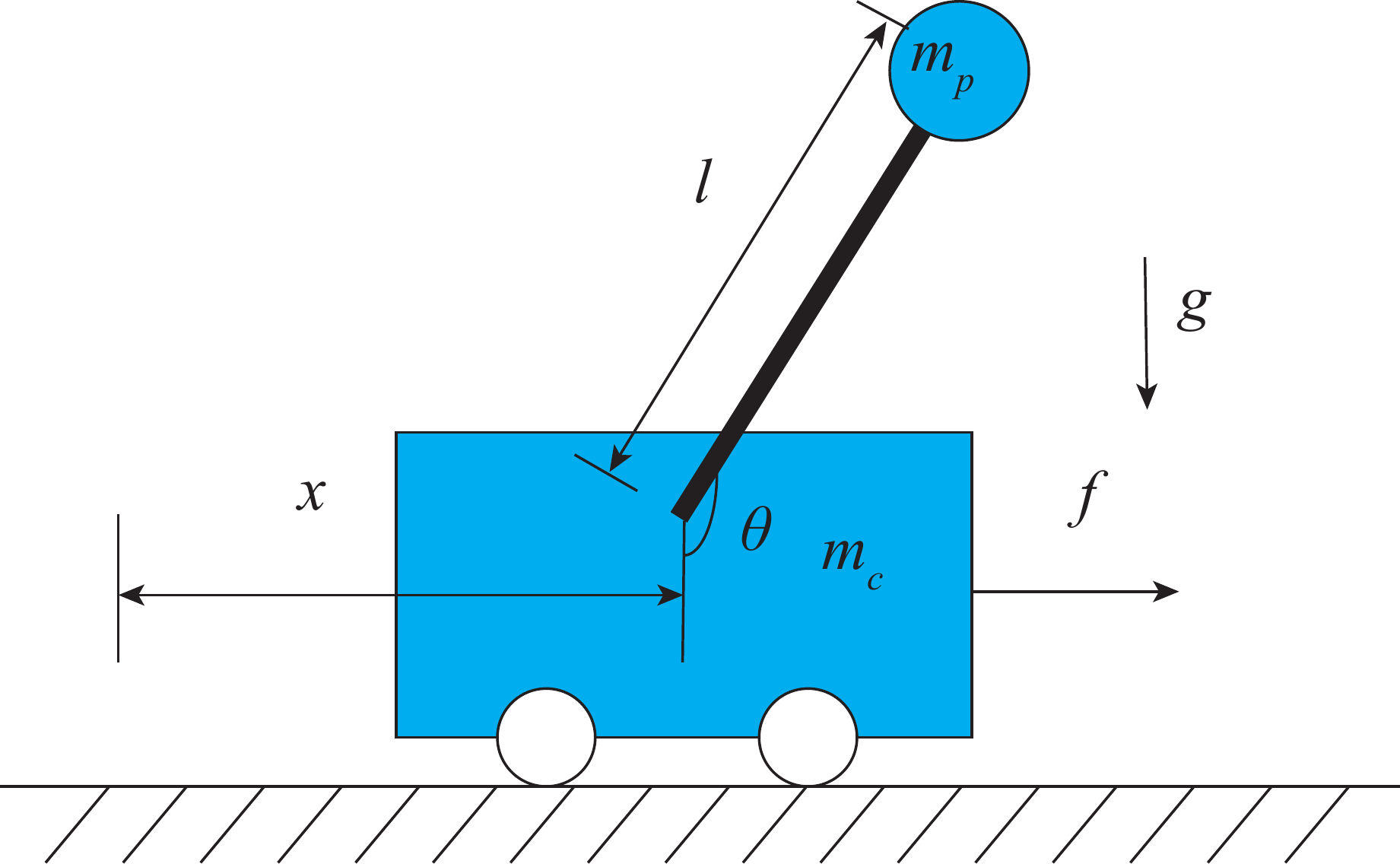}
    \caption{The cart-pole system.}
    \label{fig:cartpole}
\end{figure}

\begin{figure}[h]
\includegraphics[width=0.47\linewidth]{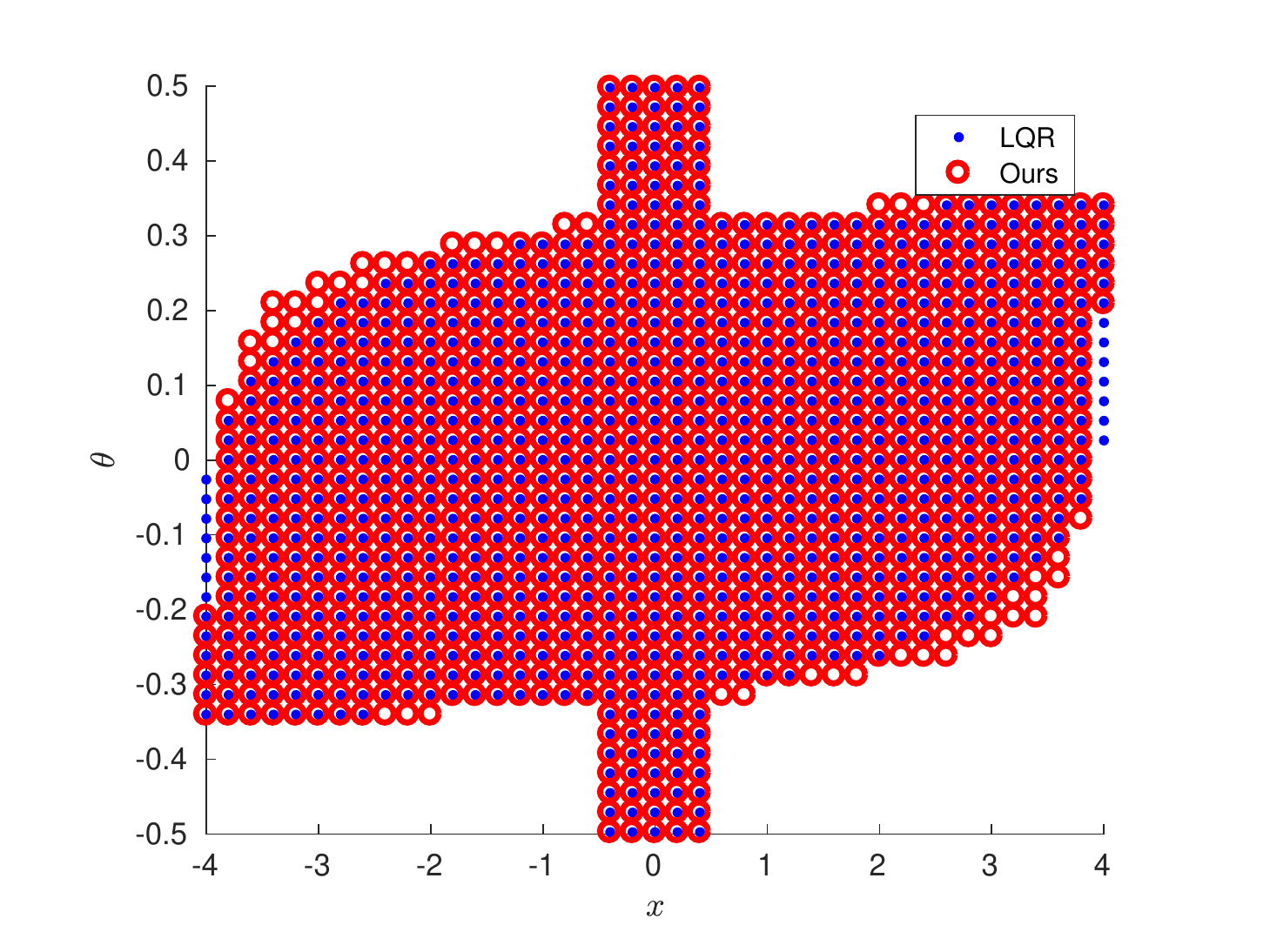}
\includegraphics[width=0.47\linewidth]{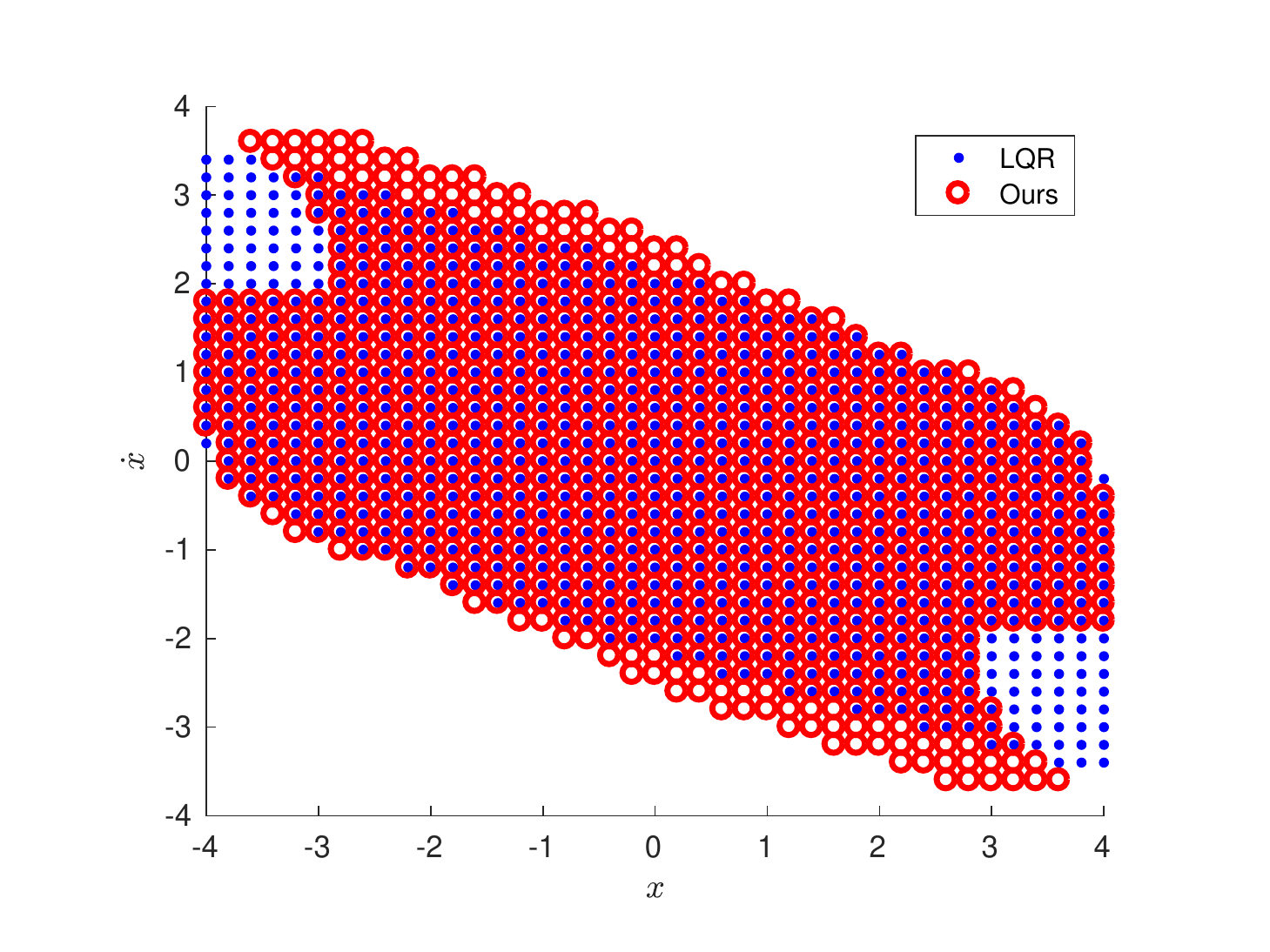}

\includegraphics[width=0.47\linewidth]{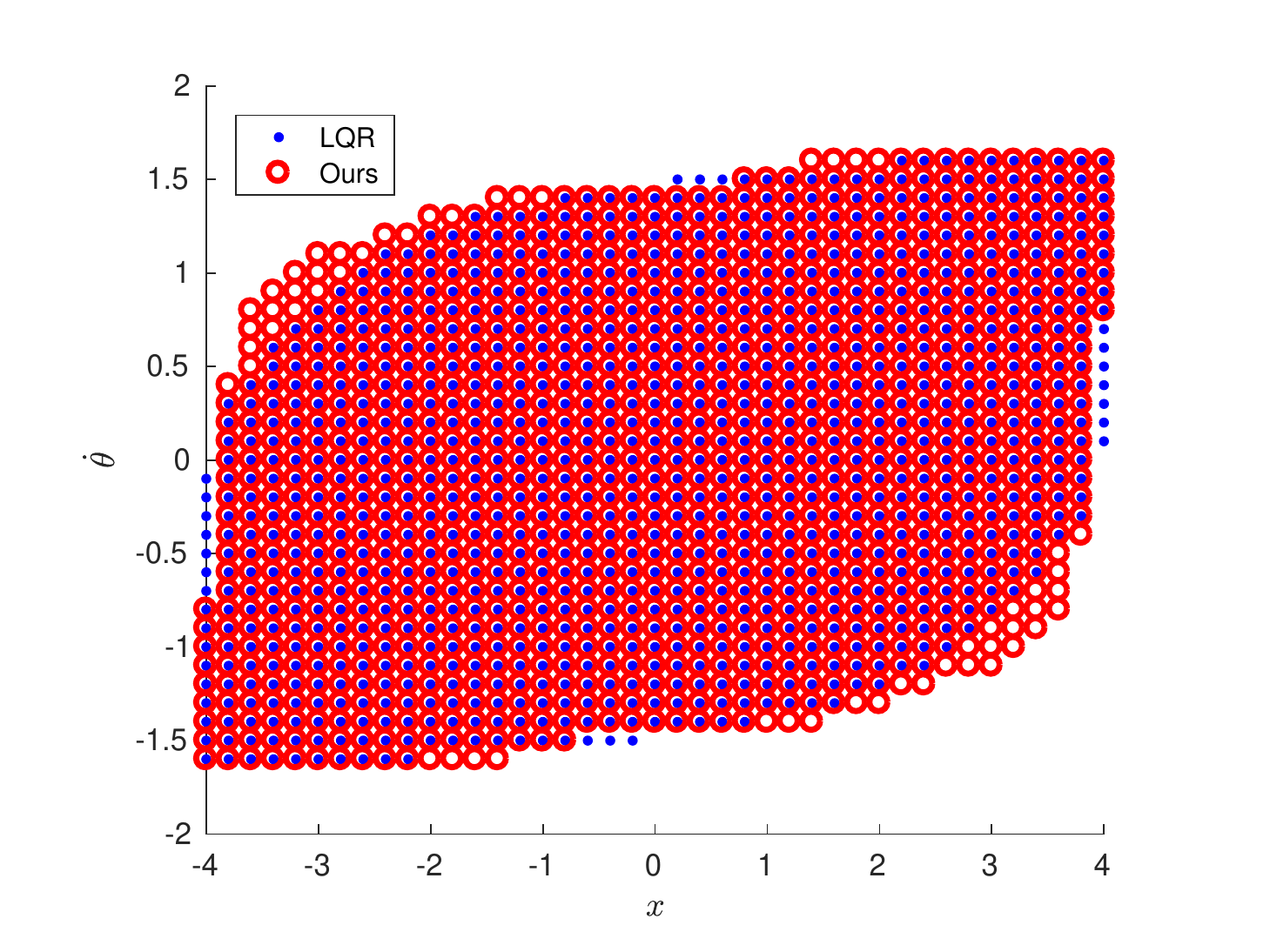}
\includegraphics[width=0.47\linewidth]{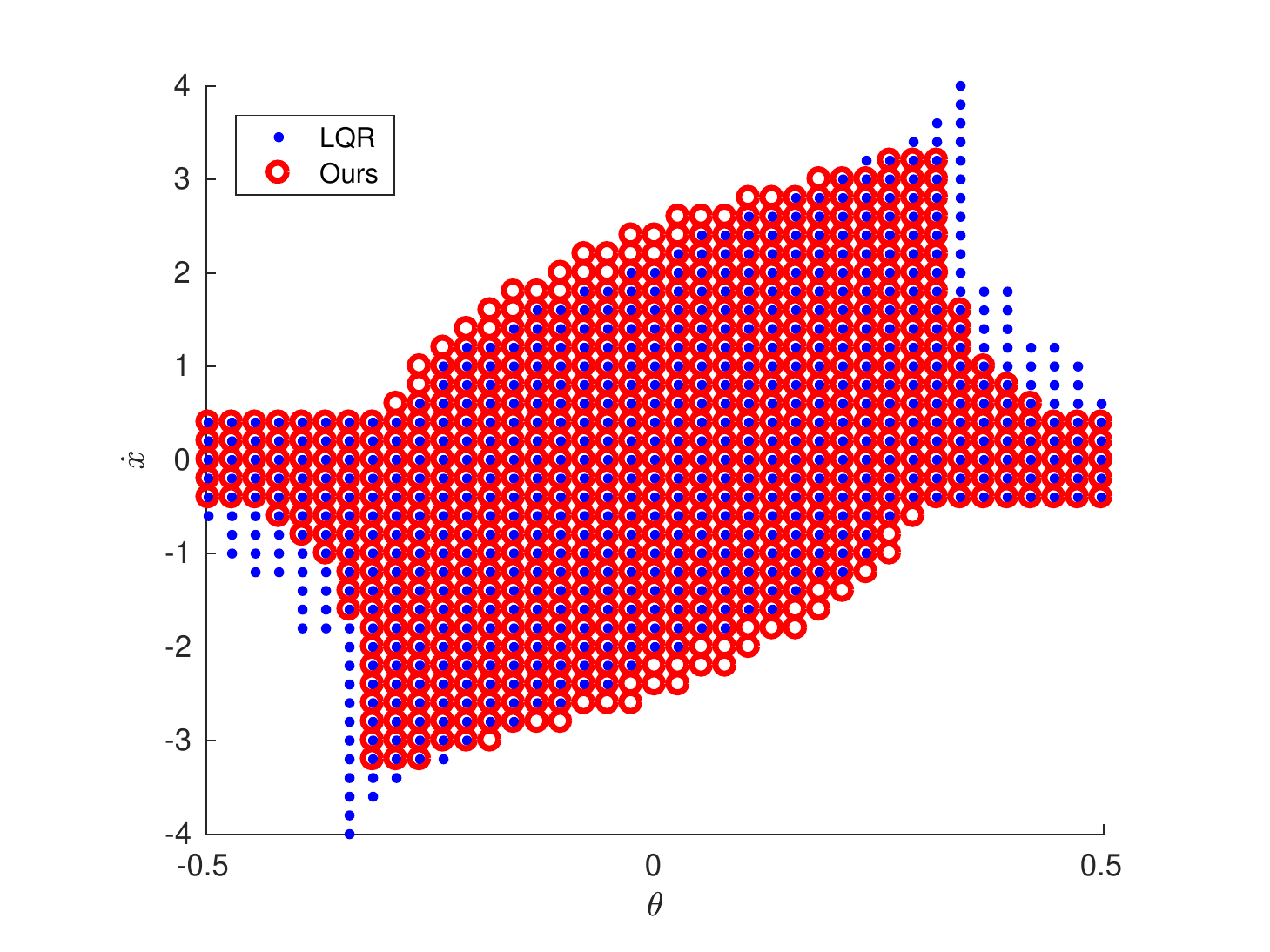}

\includegraphics[width=0.47\linewidth]{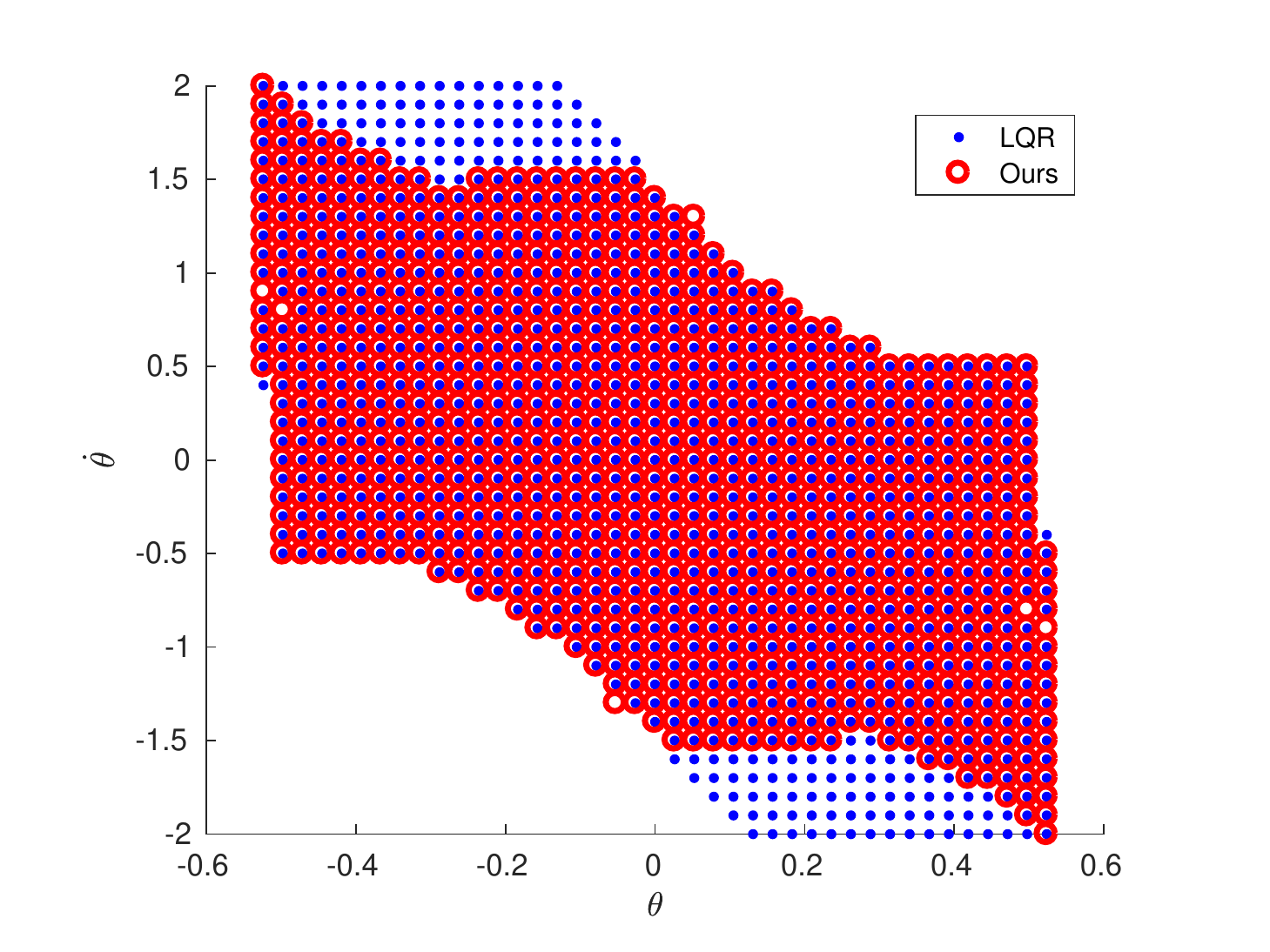}
\includegraphics[width=0.47\linewidth]{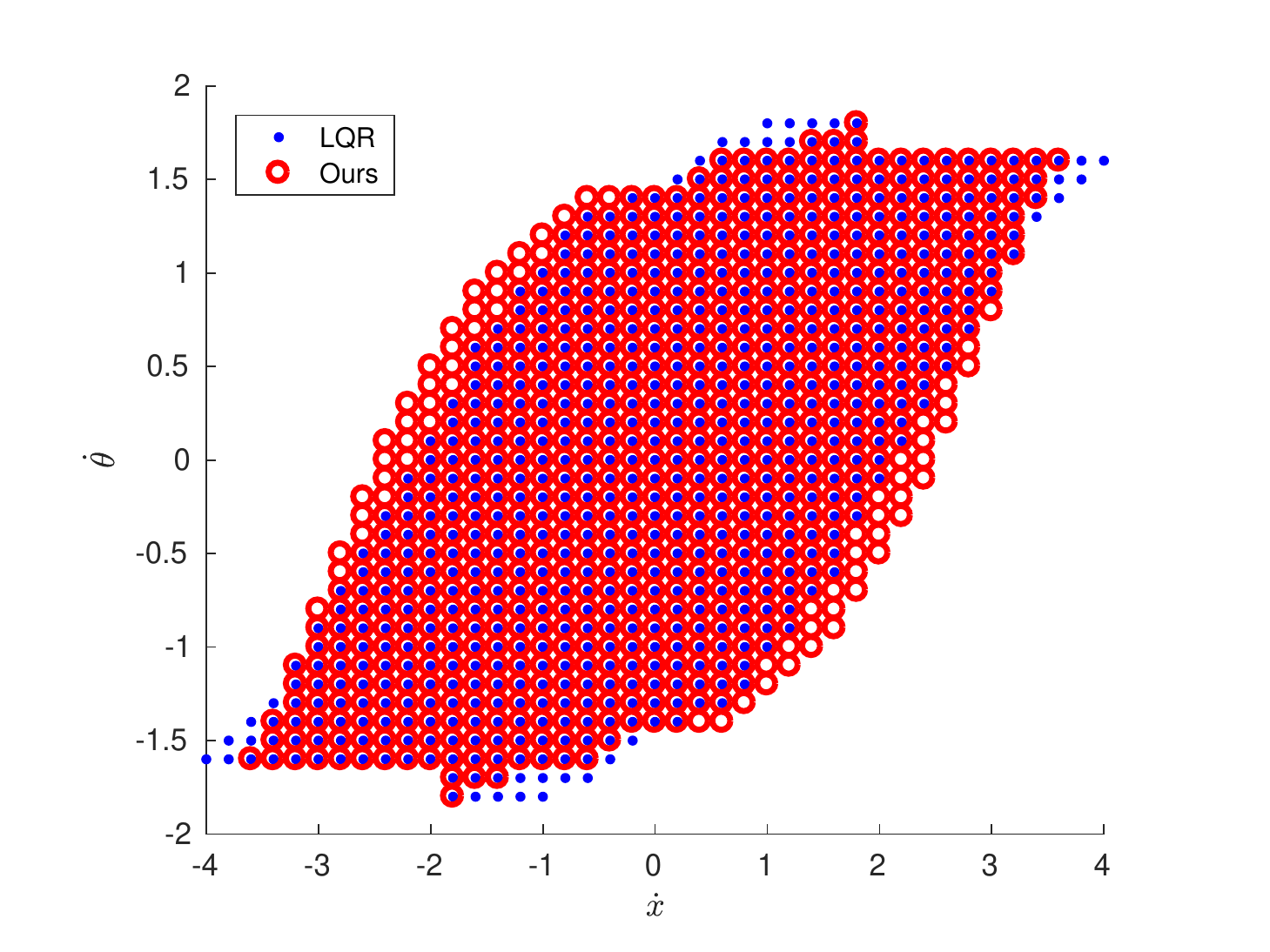}

\caption{Controllable sample points in six 2D sections. The red circles represent the controllable points using our controller, while the blue dots represent the controllable points using the infinite-horizon LQR controller.}
\label{fig:cart_pole_controllable_points}
\end{figure}

Consider balancing the cart-pole system \cite{tedrake2014underactuated}, shown in Figure \ref{fig:cartpole}, to its upright position, an unstable equilibrium.
We are allowed to apply only horizontal force on the cart, so the system is underactuated.
The equations of motion are given by
\begin{align*}
    (m_c+m_p)\ddot{x} +m_pl\ddot{\theta}\cos{\theta}-m_pl\dot{\theta}^2\sin{\theta} = f,\\
    m_pl\ddot{x}\cos{\theta}+m_pl^2\ddot{\theta}+m_pgl\sin{\theta} = 0.
\end{align*}
Let $\mathbf{x} = [x,\theta,\dot{x},\dot{\theta}]^\top$ and $\mathbf{u} = f$.
Choose $m_c = 10, m_p = 1, l = 0.5, g = 9.81$, $X = \{\mathbf{x}\in \R^4: |x| \leq 4, |\theta|\leq \pi/6, |\dot{x}|\leq 4, |\dot{\theta}| \leq 2\}$, $f\in [-40,40]$, and $Z = \{\mathbf{x}\in \R^4: ||\mathbf{x}||_\infty \leq 0.5\}$.
We Taylor-expand the equation of motion to the third order around the unstable equilibrium $\mathbf{x} = [0,\pi,0,0]^\top$, and synthesize a third degree polynomial controller.
We sample points uniformly in six 2D sections and compute the controllable points under our controller (represented by red circles in Fig \ref{fig:cart_pole_controllable_points}) using the true equations of motion. 
Each section is obtained by setting two variables to be 0.
For example, the section in the $x-\theta$ plane is $\{\mathbf{x}\in X: \dot{x} = 0, \dot{\theta} = 0\}$.
As a comparison, we also compute the controllable points under the infinite-horizon LQR controller (represented by blue dots in Fig \ref{fig:cart_pole_controllable_points}) with $Q$ and $R$ being identity matrices.

\section{Conclusion}
We have presented a controller synthesis method for discrete-time polynomial systems via the occupation measure approach.
We have also showed how to over approximate the backward reachable set of a discrete-time autonomous polynomial system and the backward controllable set of a discrete-time polynomial system under state feedback control laws.
The advantage of our approach is that we solve convex optimization problems instead of generally non-convex problems, and the computational complexity is polynomial in the state and input dimensions.
However, for controller synthesis, our method is heuristic -- stability is not guaranteed in any region.
In our future work, we will consider the discrete-time hybrid systems.

\section*{ACKNOWLEDGMENT}
This work was supported by Air Force/Lincoln Laboratory Award No. 7000374874 and Army Research Office Award No. W911NF-15-1-0166.


\bibliographystyle{plain}
\bibliography{references}

\end{document}